# ANYBODY BUT HUBBLE!

*Virginia Trimble*

*Department of Physics and Astronomy*

*University of California, Irvine, CA 92697-4575*

*USA*

**ABSTRACT**

The recent literature of history of astronomy and cosmology has included a good many suggestions for "who first recognized the expansion of the universe?" with cases having been made for Lemaître, Lundmark, de Sitter, Slipher, Shapley, Friedmann, Wirtz and perhaps others. I will touch on these but also mention others (some of whose names have not come down to us) who might reasonably be credited with some part of the basic idea, but conclude that "Hubble's Law" is the right choice ("because it was discovered by Lundmark" in accordance with Stigler's Law). Of course there are a couple of previously unsung heroes (Dose and Zöllner), and the discussion bears some traces of its origins as an after-dinner talk.

I.  INTRODUCTION - Long After Hubble

A couple of generations ago, soon after the discovery of what was then called the 3-degree, isotropic, thermal, relict radiation (CMB today) the Caltech astronomy graduate students looked at the equivalent of Fig. 1 (from Trimble 1996), drew a straight line through the values of the Hubble constant until then, and concluded it would soon go through zero and the universe start to contract. This obviously did not happen, and I now think that the straight dashed line was the wrong one to draw, since very few things are monotonic forever. Rather the historical trend should look more like the dashed curve in Fig. 1, with $H_o(t)$ leveling off to some number that will remain a consensus for some considerable time. This does seem to have happened as the input from WMAP, HST, and all have piled up (Freedman et al. 2001, Komatsu et al. 2009).

Similarly, de Vaucouleurs (1970) plotted the best estimates of the age of the cosmos and of its average density vs. time, from Copernican numbers to his (1970) present and drew straight lines through those. His conclusion was that, after centuries of change, it was unlikely that scientists had finally converged on the right answers just as he was drawing his graphs. This conclusion led him to say that the real universe must be both fractal (of lower density) and somehow much older than 10-20 Gyr. If, on the other hand, you think of a curve, with the age of the cosmos having been a human time scale for centuries, increasing rapidly with work by Kelvin, (Lindley, 2004), Jeans (eg 1934 where he argues for a cosmic age of $10^{12}$ yr), and others, and then leveling off to about a Hubble time, this is perhaps a better picture of how people's thinking really changed. Fractals apparently also level off at the supercluster scale of 100 Mpc or thereabouts, with the cosmic matter density about 1/4 of closure. It is conceivable that ages may again increase and densities decrease if multiverse, brane theory, or another of the newish ideas gains traction. If fractal structure continued out to the Hubble radius or beyond the true average density of the universe could be arbitrarily small. Some brane universes could be arbitrarily old, though we would have little or no information about cycles before our own. I have already voted against the first (Jones et al. 2004) and of necessity leave the latter to more expert opinions.

Examination of the history of other quantifiable concepts should perhaps also be described as stasis, rapid rise (or fall) and leveling off. Sizes of telescopes of a given design and numbers of sprocket-wickets sold display similar S-shaped behavior. I suspect this is also true for winning Marathon times through history. These dropped rapidly from about 3 hours at the first few modern Olympic games, and have now leveled off at 2 hr. 10 min. or so. What does not seem to have been recorded is the time it took that Greek chap running the first marathon to get back to Athens, (no Pheidippides is the one who ran from Athens to Sparta to ask for help; much further and took most of 2 days). We can only hope that it will also be true for a number of human beings, numbers of other species wiped out, and so forth.

II. LONG BEFORE HUBBLE

For once, we don't start with the ancient Greeks, but with some other cultures whose creation myths involved expansion (Leeming & Leeming 1997). Best-known is probably the Chinese Phan-ku (Pan-gu, and other spellings) who separated earth and sky by growing 10 feet per day for 18,000 years, so that heaven and earth are now separated by 90,000 li or 30,000 miles (numbers from Leeming & Leeming 1997); other sources give somewhat different numbers, and the combinations are not quite self-consistent for any choice of 1 mile = N li. The Norse Yggdrasil, an over-sized ash tree, apparently performed a similar "separation of heaven and earth" function. The only drawing I've seen on paper appears in Agatha Christie's (1946) *The Hollow* and seems to belong to the date palm family (the drawing is, of course, a False Clue). Another version appears in a Wiccan catalogue. Third, and illustrated in the nicest pictures, is the Ancient Egyptian air god Shu, supporting the sky goddess Nuit after raising her above the earth god Geb.

By no means all ancient creation myths are of this type. Leeming & Leeming (1997) list at least 16, a few of which would seem to have modern analogies, for instance Creation from a Cosmic Egg (Lemaître's primeval atom, eg. 1950), Creation from Ancestors (brane worlds, Vilenkin 2006), Creation by Emergence (E.O. Wilson 1998). Creation from Chaos and Creation by Word are the generic types of the old and new testament. There tend also to be a fair number of turtles involved. Some of them trigger the growth or expansion of something else (Maidu creation), but it is not clear that any culture ever took seriously the lovely old wood-cut picture of the earth as the back of a giant tortoise, swimming in a bowl of water, held on the back of an elephant, standing on another tortoise, and from then on "turtles all the way down".

A later question in the history of the cosmos is "what holds up the sky?", and there have also been many answers to that (Trimble, Martinez, & Hockey 2013).

III. EINSTEIN MISSES A BET

We have been told umpteen godzillion times that Einstein thought the universe had to be stable or static because there was no observational evidence for expansion or contraction. In fact, Campbell (1911) at Lick had identified and measured what he called the K-term, meaning that, averaged over the sky, the radial velocities of stars were not zero, but something like +3km/sec. Many of the stars contributing to this were of type B and 300-350 pc away (Russell et al. 1927 pp. 879).

Now, 325 pc divided by 3 km/sec is just about $10^8$ years, a good age for a slightly post-Kelvin cosmos. But, as many others have remarked, Einstein was not very interested in astronomical observations, (including Slipher's redshifts), when he first heard about, and rejected, Friedmann's expanding solution to his equations. As you have also often heard, the rejection was based, first, upon an Einsteinian mathematical error, and then, when he realized that Friedmann's math was right, on the feeling that the physics was awful.

The planned talk at this stage had pictures of Pan-ku, Geb-Shu-Nuit, a particularly shrunken 6th century "Christian Topography" from Cosmas Indicopleustes (Fig. 1.8 from North 2008), and what is probably the first picture of the Milky Way to show the globular clusters grouped far from the solar system published in 1909 by Karl Bohlin, director of Stockholm Observatory. His computed direction to the galactic center was correct (H. Abt in Hockey et al. 2007, Bohlin (1909)).

IV. CLEARING OUT SOME UNDERBRUSH

First, here are some questions I will not try to answer: (1) Who discovered the expansion of the universe? (2) Who discovered Hubble's Law? (3) Should Hubble's law be renamed??

If the primary question was "Who discovered Hubble's law?" the primary contenders are probably Hubble and Lemaître. In that context Block (2011), Shaviv (2011), Livio (2011), and van den Bergh (2011) have relevant things to say.

Milton Lasseell Humason was part of the story a little later. He was a remarkable person (for which see several passages in Sandage 2004), whose often multi-night spectrograms yielded most of the redshifts used by Hubble after the well-known 1929 paper, which had Slipher redshifts. Just one Humason story, courtesy of Uncle Allan:

"Why the first week in November never found me in Pasadena remained a mystery until Nicolson revealed that it had been no scheduling quirk. Humason's annual contribution to the Republican party was to arrange (in his capacity as chief telescope-scheduler) for every Democrat on the nighttime-observing staff to be high atop the mountain – nowhere near the polls – on Election Day."

Second, some questions to which I will attempt partial answers: (1) Who wrote which words, when? (2) Who measured what (redshifts, angular diameters, apparent magnitudes, surface brightnesses) when? (3) Who wrote down which equations, when? (4) Who put which numbers forward from those equations, when? (5) Who plotted which graphs, when? (6) Who tabulated numbers (etc.) that could have been plotted to yield interesting results, but did not, when (Lundmark & Wirtz will appear in this context).

The major secondary sources used were Kragh (1996), North (2008), Sandage (2004), Nussbaumer & Bieri (2009, 2011), Bertotti et al. (1990), Martinez et al. (2001), and a number of very informative e-discussions with Sidney van den Bergh, and his chapter in Livio et al. (1997).

V. MODERN PRE-HISTORY

The underlying idea, which went back to the first astronomers to have a few proper motions in hand, was to measure the motion of the Sun (later the Galaxy or the Universe) relative to some group of stars or other astronomical objects. That the Sun, locally, is moving at 20 or 30 km/sec in the general direction of Hercules, relative to nearby stars, has been a very stable result. Campbell's (1911) K-term for moderately distant stars was apparently some combination of slightly erroneous laboratory wavelengths and a bit of actual expansion of the B stars in Gould's Belt.

The first spectrogram of M31 came from Julius Scheiner at Potsdam in 1899, and he reported that there were absorption features rather like those in the solar spectrum.  He also notes (Scheiner 1899) that an integrated spectrum of the Milky Way would look more like Type I stars (like Vega with dominant hydrogen) than like Type II (like the sun, and the core of M31).  Edward Fath, working first with the Crossley at Lick (1906-09) and then at Mt. Wilson (1909-12) recorded a number of other absorption, stellar-like spectra for other spiral nebulae, but (unfortunately) also had NCC 1068 and one other Seyfert galaxy in his sample, showing strong, centrally-condensed, emission lines to confuse the issue (R.P. Lindner in Hockey et al. 2007 who cites Fath (1909, 1911, 1913)).

Slipher's 1912 (December 3-4) spectrogram is fuzzily reproduced in Russell et al. (1927, p. 848). Definitely the F, G, H and K absorption features are there.  And H.D. Babcock's (1939)  rotation curve for M31 made use of emission features from nebulae out to about 30 kpc from the center.  His rotation curve was nearly flat, and was recognized as part of the history of dark matter only considerably later. Its reception by the contemporary community is one of the reason you now remember Babcock as a solar physicist.

VI.  PEOPLE AND PAGE NUMBERS

This is an attempt to list, in roughly chronological order, as many as possible of the papers prior to Hubble (1929) that you might want to consult on the topic of early history of the expanding universe. A few words about the contents of each follow the reference; more is said about some of them in the following sections.  In the intended oral presentation, the list would have been interleaved with pictures of Slipher, Lundmark (whom, Nina Strömgren Allen said to the present author, her parents were somewhat afraid of).  Shapley, Strömberg, Lemaître, Wirtz, and August Dose.

1907.   J. Holetschek. Ann. Wein. Stern. Total apparent magnitudes and angular diameters of a number of spiral nebulae
1915.    Vesto Melvin Slipher.  Pop. Ast. 23, 21. 15 line shifts; up to 41 by 1922
1916.   G.F. Paddock.  PASP 28, 109.  Hoped excess of redshifts would go away with further data, especially from other side of sky
1916.   O.H. Truman. Pop. Ast. 24, 111; net solar motion toward RA = 20$^h$ Decl. = - 20º at 670 km/sec

1916   R.K. Young & W.E. Harper. JRASC 10, 134. "velocity of universe" (meaning Galaxy) toward RA = 24$^h$, Decl. = - 12$^o$, at 598 km/sec

1916.   Willem de Sitter MN 76, 699; 1917 MN 78, 3. de Sitter solution in which redshifts might plausibly occur, perhaps quadratical in distance; much more widely known than Friedmann & Lemaître solutions, so that people looked for his effect. De Sitter space has constant negative curvature and repulsive cosmological constant. Anti-de Sitter space has constant positive curvature coming from an attractive cosmological constant and has applications in supersymmetry and string theory.

1918.   Carl Wilhelm Wirtz AN 206, 109; also 1922, 1924, 1925 Scientia 38, 303. positive velocities mean expansion of the system (more below)

1919.   Knut E. Lundmark AN 209, 369. M31 at 220,000 pc

1919.   Harlow & Martha Betz Shapley. ApJ 50, 107 "the speed of the spiral nebulae is dependent to some extend upon apparent brightness, indicating the relation of speed to distance or possibly to mass." which in 1929 he claimed had been a decade's anticipation of Hubble (nonsense, said Sandage, 2004 p. 503).

1921.   J. Hopman AN 214, 425, angular diameters of spirals

1922.   Ernst Öpik. M31 at 450 kpc, ApJ. 55, 406

1922.   Kornel (Cornelius) Lanczos. Ph. Zs. 2, 539, theoretical distractor (in the sense of wrong answers in a multiple choice test)

1922.   Alexander A. Friedmann Zs f Ph 10, 377 and (1924) 21, 329. Multiple solutions to Einstein equations, including expansion from singularity

1923.   Hermann Weyl. Ph Zs 24, 130 & 230, implicitly that de Sitter solution implies linear velocity-distance relation, but otherwise a distractor (see Misner et al, 1970 p. 758)

1924.   Ludvik Silberstein MN 84, 363, attempt at velocity-distance, but using both negative and positive velocities to include globular clusters

1924.   Knut Lundmark, MN 84, 247, & (1925) MN 85, 865. also considered negative velocities and globular clusters, but case can be made that he came very close to Hubble's Law

1925.   Gustave Strömberg ApJ 61, 353, solar motion relative to many different stellar populations, including globular clusters (which do not suggest any velocity-distance relation)

1927.   Georges H.P. Lemaître. Ann. Soc. Sci. Bruselles 47, 49. Both an expanding solution of the Einstein equations and examination of Slipher velocities, yielding a possible Lemaître constant of about 600 km/sec/Mpc; also possibility of linear velocity-distance in de Sitter space; but his MIT PhD dissertation was on the Tolman-Oppenheimer-Volkoff equation (before TOV)

1927.   August Dose. AN 229, 157 (a student of Wirtz) velocity-distance correlated, not obviously linear; the first paper in German of which I've read every word in a long time, triggered by the deaths of three German-speaking collaborators in less than a year – Harry Lustig, Meinhard Mayer, and Hilmar Duerbeck)

1928. Howard Percy Robertson. Phil. Mag. 5 895, very similar to Lemaître, predicting a linear velocity-distance relation (independently); the line element retains his name as the Robertson-Walker metric

1929. Edwin Powell Hubble. PNAS 15, 168, circumspexi, and phrases like "Hubble ratio," "Hubble's law" and "Hubble's velocity-distance relation" are widespread in the literature by 1933 (Trimble 2012). Richard Chase Tolman said important things, but only post-Hubble

A few words about units and magnitudes:
Wirtz, Dose, and sometimes Lundmark used distances in units of d to M31 (for which there was about a factor of two difference between Öpik and Lundmark). NGC 224 = M31 in case you had forgotten.

Distances from apparent (blue) magnitudes: If $M_B$ = -15.2 on average (Hubble's number) yields a Hubble constant of 526 or so, then $M_B$ = - 19 (the peak of Wirtz's luminosity distribution) will yield H = 69 or thereabouts.

If distance is to come from apparent diameters plus some standard size, then one must use $1/\Theta$ to get a linear velocity-distance curve. The choice of $-\log(D_m)$, with $D_m$ as observed in arc minutes and a calibration with M31 then turns the relation into one that at least goes the right direction.

VII. SO THEY SAID, WROTE, OR DREW, A FEW LEMAÎTRE ANECDOTES, AND WIRTZ

<u>De Sitter</u>, 1917 in MN. "Spiral nebulae are probably amongst the most distant objects we know. There are 3 with good velocities - two recession and one approaching...maybe it is a hint." The questions of how many redshifts are enough, and whether you can make it up on quantity if you lose money on every sale were repeatedly raised in the early days of observational cosmology (and indeed onward to the steady state vs. big bang days).

<u>(Shapley)</u>[2] in 1919 "The speed of spiral nebulae is dependent to some extent upon apparent brightness, indicating a relation of speed to distance or, possibly to mass." Their apparent magnitudes were from Holetschek. Since for Shapley at this time, the Milky Way was the entire universe, no cosmological conclusions can be drawn from his choice that 6 brightest S's have a mean velocity of +49 km/sec, while the other 19 average 726 km/sec. Let his spirit rest content with the enormous achievement of having gotten us out of the center of the Milky Way, and being the father of the 2012 Nobel Prize winner in Economics, Lloyd Shapley!

<u>Lundmark</u> 1924 plotted velocity vs. distance (in Andromeda units); got some kind of relation, but "not a very definite one." The plot implies a recession velocity of 15-30 km/sec per Andromeda distance, or 75-150 km/sec/Mpc with M31 at 220 kpc, a very strong pro-Lundmark point notes Steen (2011, 2012, 2013). But Lundmark's own Andromeda distance was 36 kpc, so his "H" = 550 km/sec/Mpc or thereabouts. He also attempted to add a quadratic term, so that $K = +512 + 10.365r - 0.047 r^2$, where r is distance in Andromeda units. That distance, he said is 30 times the diameter of the Milky Way, from his own work and that of Charlier. Then, ignoring the $r^2$ term, some $H = dK/dr$, and if the diameter of the Milky Way is 6 kpc (Kapteyn's number), H = 57.6 km/sec/Mpc.

<u>Lemaître</u> didn't actually plot his distances, but only tabulated them, 0 - 4 Mpc, with velocities 0 to 2000 km/sec. Seitter and Duerbeck (1990) turned these into a plot, and indeed (no surprises) H = 575 km/sec/Mpc puts half the dots on each side of the line. He was, of course, a priest, required to say mass every day, but not necessarily to listen to anyone else say it. Thus, when he visited the Royal Observatory at the Cape and they laid on a car to take him to the cathedral each day, he declined its use. And when invited by the local bishop to preach there, he said he would be happy to, and would like to talk about the expanding universe, at which point the invitation was withdrawn (story courtesy of Michael W. Feast).

<u>Lemaître</u> did not attempt to claim additional credit for his early work. His teaching was chaotic, and his later publications, like his early ones, were in journals with low impact factors. He had intended priesthood even before the horrible experiences of WWI (in which he served, won a medal, and saw the wreckage of Louvain). As a founding member of the Pontifical Academy of Sciences, and its president 1960-66, he advised the Pope not to link physical cosmology with creation (information courtesy of Christoffe Waelkens at Louvain). General Ludendorf (commander of the Germans in Belgium) had a brother who was a spectroscopist at Potsdam and who appears briefly in Hockey et al. (2007) (information courtesy of George Wallerstein).

How did Lemaître find time to say his holy offices when he attended very intense conferences? William A. Fowler brought back from the 1957 Vatican meeting on galaxies the following Lemaître response when asked: Oh, I just wait until after breakfast, when the ladies say they are going upstairs to get dressed and will be back in just a moment. This gives me plenty of time. Those ladies were, of course, what we now call "accompanying persons."

All is, naturally, more obvious with 20-20 hindsight. Van den Bergh (in Livio et al. 1997) plotted the 41 velocities Slipher had provided by 1922 vs. apparent B magnitudes (ignoring the 3 negative numbers, which were actually the 3 brightest galaxies). A bisector line, or one tilted down a bit to allow for some Malmquist bias, would give a value of H to ± 10%, like all values of H from ancient times to the present, but with an absolute value that depends entirely on what you think the average brightness of a spiral is! If $M_B = -19$ (as per Wirtz), then H = 100. If $M_B = 15.2$ (do it yourself; you need the practice!)

Wirtz wrote in 1922 "...the most striking major process...an expansion of the system of spiral nebulae with respect to our own position...and the nearer or more massive nebulae show less expansion than the distant nebulae or those of lesser mass." He determined $V(km/sec) = 2200 - 1200 \log (D_m)$ which goes the right direction for V increasing with distance, but does not suggest a value for H if you take dV/dr, since $D_m$ = angular diameter in arc minutes and so scales like 1/r. But 1000 km/sec at $D_m = 1$ and a luminosity distribution for galaxies peaking at $M_B = -19$, would make for an H much smaller than Hubble's.

Seitter and Duerbeck (in Bertotti et al. 1990) have taken the data from Wirtz's tables and constructed diagrams of velocity vs. apparent magnitude, velocity vs. distance for spirals, angular diameter vs surface brightness for ellipticals (which must reflect intrinsic properties since surface brightness is not distance-dependent for nearby galaxies), and the distribution of absolute magnitudes Collectively, they are "not inconsistent" with a linear velocity-distance relation, but no value of a Hubble constant could have been extracted from the data examined in Wirtz's way. They carry the story on through 1936 and also examine much of Wirtz's other work on stars and what we now call galaxies.

VIII. DOSE, HUBBLE, AND BEYOND

Just for fun, try Googling A. Dose without further disambiguation. You will learn a great deal about millimoles per milliliter, and not much about km/sec per Mpc. August Dose was actually Wirtz's student, born in 1902 in Bad Segeberg and died 1983 in Essen. He earned his PhD at Kiel in 1927, spent a couple of years at Hamburg, was in the editorial office of *Astronomsche Nachrichten* (whose current editorial office provided basic biographical information and a picture of him) from 1929 to 37, worked at the Bremen Focke-Wulf aircraft factory 1937-51, and finally was part of the Aachen Technical University of Dynamics until his 1957 retirement (Fig. 2).

His work in one sense did not represent much of an advance over that of Wirtz, as Seitter and Duerbeck (1990) noted. Perhaps most important is that he had some Mt. Wilson and Lick radial velocities as well as Slipher's from Lowell, and found that the same galaxy observed from more than one place generally had about the same $V_r$. He divided his sample into three by angular diameter, with breaks at 3' and 6'. The average Vr was indeed largest for the smallest angular diameters (most distant galaxies). The middle group had a mean $V_r$ of 719 km/sec. If these are "Kapteyn" galaxies, with diameters of 6 kpc, then H = 140 - 280 km/sec/Mpc.

Hubble's own 1929 diagram you have all seen, with its 10% uncertainty and a vertical axis said to be in km rather than km/sec. Seitter and Duerbeck remarked that Wirtz often used km for km/sec, and that German speed limit signs continued to do so down more or less to the present.

Figure 3 is my own favorite plot of this vs. that. It is values of the Hubble constant vs. poster number at the conference whose proceedings appear in Livio et al, (1997, STScI Symposium 10, May 1996), where I had the privilege of providing the concluding remarks. The correlation looks about as good as some of the early determinations of H! The next step after concluding remarks is, of course, the after-dinner talk. And beyond that comes what the late Maurice Goldhaber called the 8th age of man, "My but you're looking well!" So, I apologize that a brief, tiresome illness kept me away from the Slipher Centenary Symposium.

IX. CONCLUSIONS

The verdict of history, including citation counts, is that the velocity-distance relation should be called Hubble's law, and I am inclined to agree, for reasons given at excessive length in Trimble (2012). If, however, you ask slightly different questions, like those at the beginning of Section IV, cases can be, and have been made for:

1. Vesto Melvin Slipher (Way and Hunter, 1912)
2. William de Sitter (de Sitter 2000)
3. Georges Lemaître (Nussbaumer & Bieri 2011 and references therein; Blanchard 2001)
4. Harlow Shapley (Sandage 2004, quoting Shapley)
5. Alexander Friedmann (Belenkiy 2012)
6. Carl Wirtz (Seitter & Duerbeck 1990 and a talk at an STScI Symposium in 1995, not included in the proceedings).
7. Knut Lundmark (Steen 2011, 2012, 2013)

If you decided to ask who first suggested curved space or space-time for the universe, the answer would, of course, be "anybody but Einstein." Kragh (2012) has just called attention to Karl Friedrich Zöllner as the winner, though with various mentions of Karl Schwarzschild, Robert S. Ball, Simon Newcomb, and Paul Harzer.

Did the Zöllner universe expand? Not exactly, but he did argue that, in an unbounded and therefore infinite, Euclidean space any finite amount of matter would evaporate and dissolve to zero density in an infinity of time, from which he deduced that space or time(or both) must be finite. As for why astronomical contemporaries of Zöllner did not consider the consequences of space being non-Euclidean, "they had no need for the hypothesis." It is left as an exercise for the reader to recall what famous quote this echoes.

ACKNOWLEDGEMENTS

I am indebted to organizers for the original invitation to participate in the Slipher Centenary. Alison Lara keyboarded the text with her usual expertise, for which I am most grateful.

FIGURE CAPTIONS

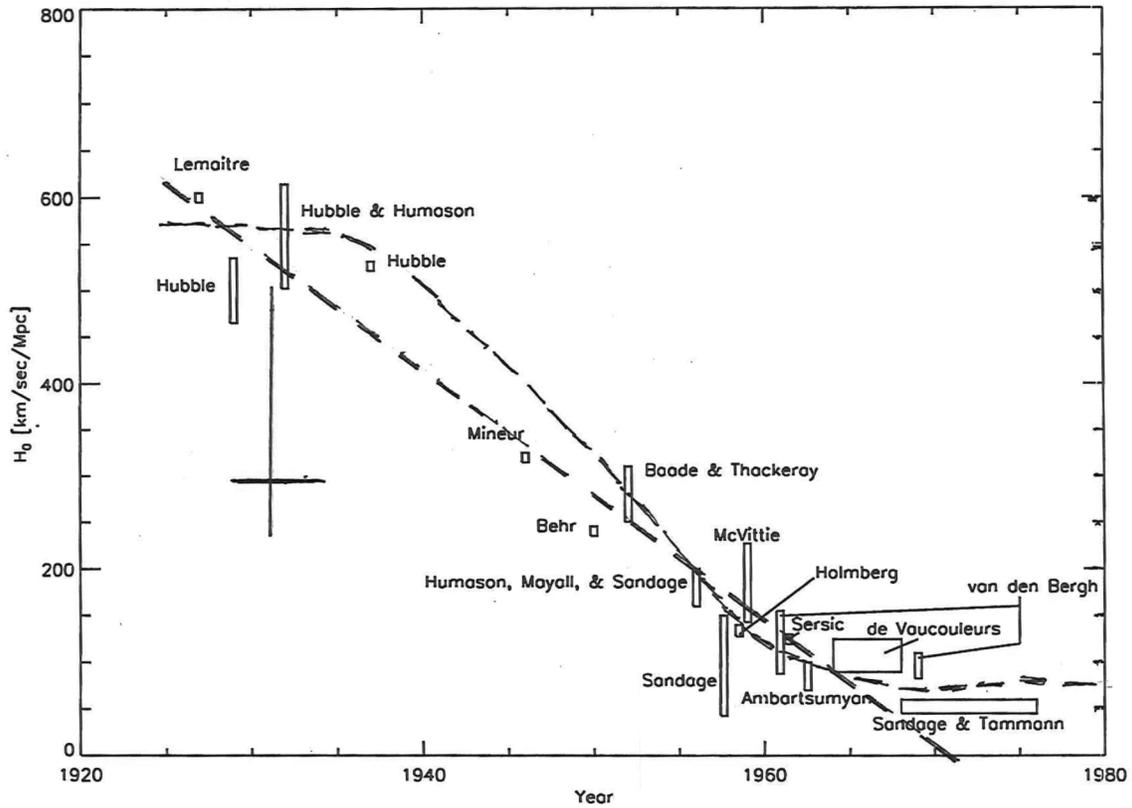

**Fig. 1:** Plot of published values of the Hubble constant through time from 1927 to 1977. The straight line is probably the wrong way to look at things (and, of course, predicted H = 0 by 1975). The curved line, with a period of stasis, fairly rapid change, and then convergence on some number near 75 is a better bet. The inverted cross comprises numbers implied by work from Jan Oort in the 1930s. Other very approximate points that could be tucked in include 460 km/sec/Mpc from H.P. Robertson in 1928, 450 ± 50 from W. de Sitter in 1930, and 571 from F. Zwicky in 1933.

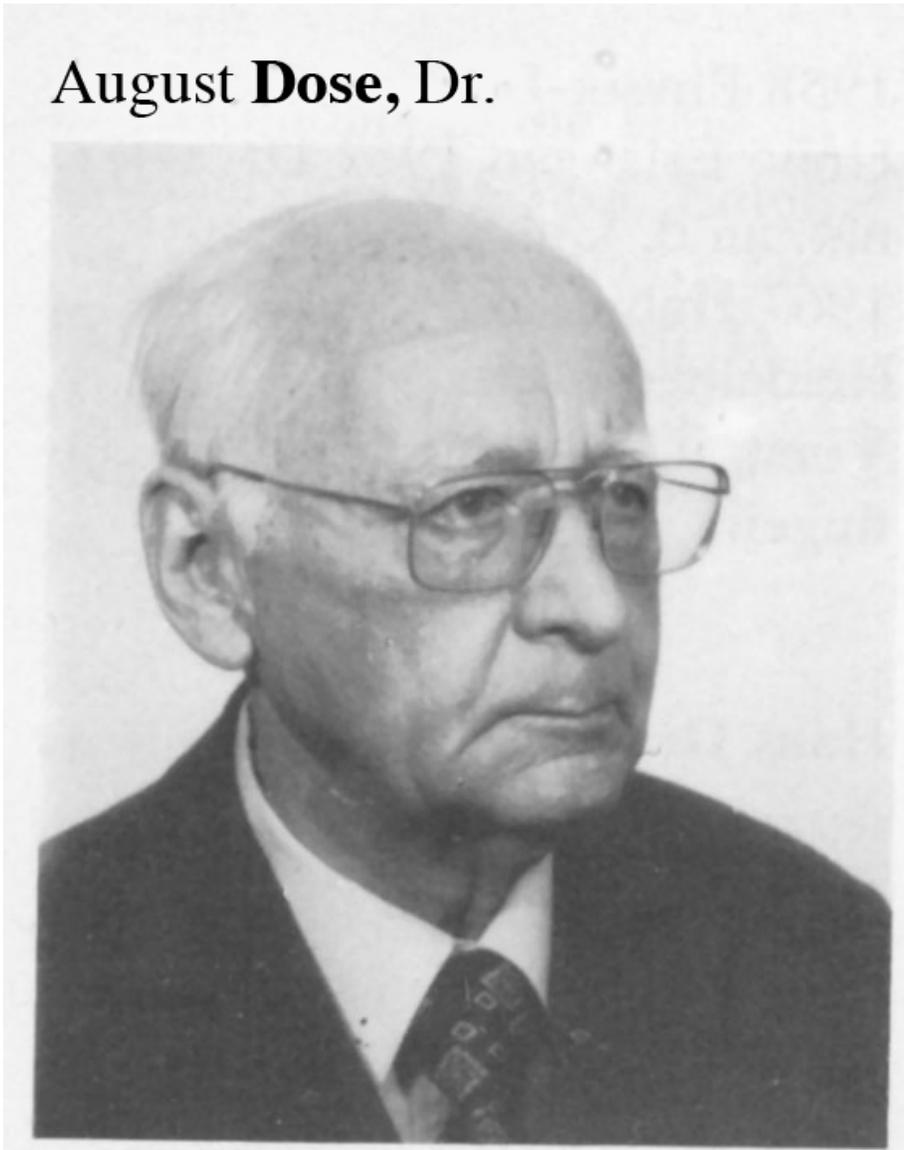

**Fig. 2:** August Dose (1902 – 1983). No, he did not pre-discover Hubble's Law in his 1927 thesis (Zur Statistik der nichtgalaktischen Nebel auf Grund der Konigshtuhl-Nebellisten mit einer Bemerkung ber die Radialbewegungen der Spiralnebel), but he did check that Mr. Wilson and Lick Observatory radial velocities agreed with the Lowell ones. Courtesy: Astronomische Nachrichten, in whose offices Dose worked 1929 – 37.

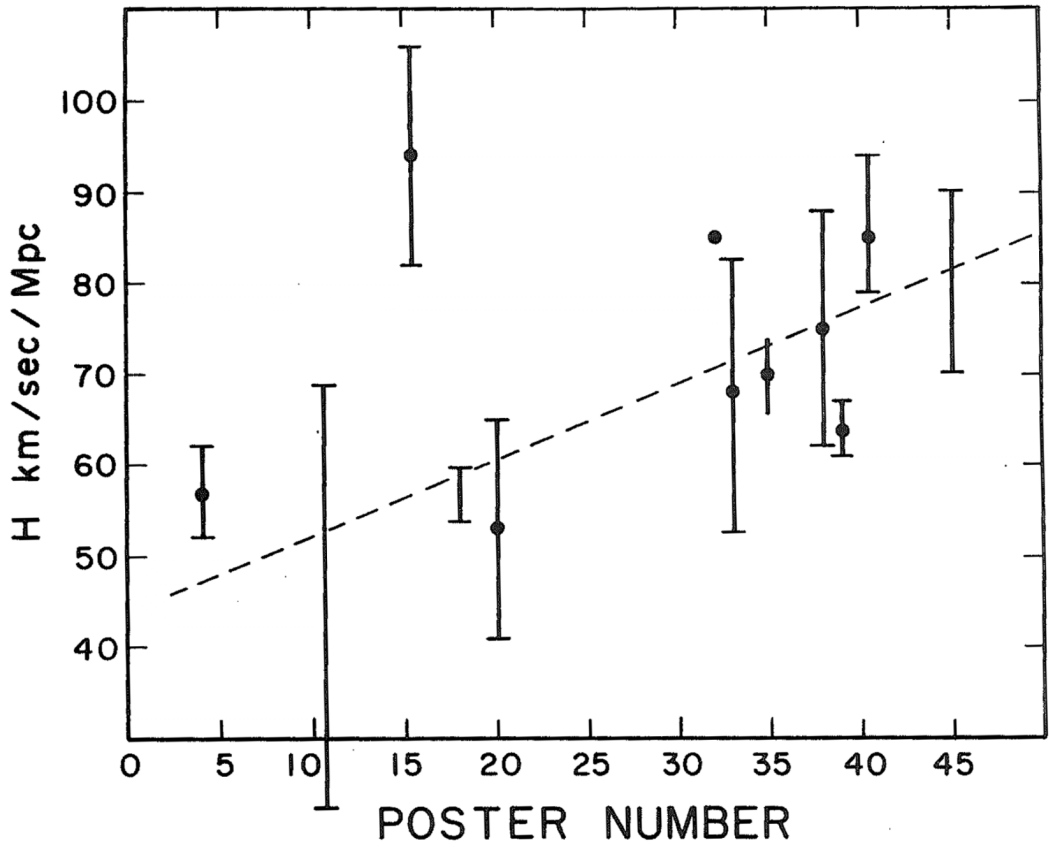

**Fig. 3:** Plot of values of the Hubble constant suggested in the poster papers at a 1995 conference held at Space Telescope Science Institute, vs. poster number, that is, an alphabetical ordering of the first authors. The correlation is about as good as some of the others in cosmology.